\documentstyle[twocolumn]{article}
\begin{document}

\twocolumn[\hsize\textwidth\columnwidth\hsize\csname@twocolumnfalse\endcsname

\title{ Quantum key distribution relied on trusted information center}
\author {$^1$Guihua Zeng, $^2$Zhongyang Wang, and $^1$Xinmei Wang \\
$^1$National Key Lab. on ISDN, XiDian University, Xi'an 710071, China\\
$^2$Shanghai Institute of Optics and Fine mechanics, Academia Sinica,\\ P.O.Box 800-211, Shanghai 201800, China }

\date{}
\maketitle
\begin{abstract}

Quantum correlation between two particles and among three 
particles show nonclassic properties that can be used for providing 
secure transmission of information. In this paper, we propose two 
quantum key distribution schemes for quantum cryptographic network, 
which use the correlation
properties of two and three particles. One is implemented by the
Greenberger-Horne-Zeilinger state, and another is implemented by 
the Bell states. These schemes need a trusted information center 
like that in the classic cryptography. The optimal efficiency of the proposed 
protocols are higher than that in the previous schemes. 
\vspace{1.0cm}
\end{abstract}
] 

\begin{flushleft}
{I. Introduction}
\end{flushleft}

Since the first finding that quantum effects may protect privacy information transmitted 
in an open quantum channel by S.Wiesner [1], and then  by C.H.Bennett and G.Brassard [2],
a remarkable surge of interest in the international scientific and industrial
community has propelled quantum cryptography into mainstream computer science
and physics. Furthermore, quantum cryptography is becoming increasingly
practical at a fast pace. Quantum cryptography is a field that combines quantum theory with 
information theory. The goal of this field is to use the law of physics to provide secure 
information exchange, in contrast to classical methods based on (unproven) complexity 
assumption. Current investigations of quantum cryptography involve three aspects: quantum key distribution (QKD) [3-19], 
quantum secret sharing [20,21], and quantum bit commitment and its application [22-23].
In particular, quantum key distribution became especially important due to technological 
advances which allow their implementation in laboratory.
The first quantum key distribution prototype, working over a distance 
of 32 centimeters in 1989, was implemented by means of laser transmitting in free space [9].
Soon, experimental demonstrations by optical fibber were set up [10]. 
Now the transmission distance is extended to more than 30Km in the fiber [11], and more than 205m in the free space [12].

Quantum key distribution
is defined as a procedure allowing legitimate two (multi-) users of communication channel
to establish exact two (multi-) copies, one copy for each user, of a random and secret
sequence of bits. Quantum key distribution employs quantum phenomena such as 
the Heisenberg uncertainty principle and the quantum
correlation to protect distributions of cryptographic keys. 
QKD is a technique that permits two (or multi-)
parties, who share no secret information initially, to communicate over an
open channel and to establish between themselves a shared secret sequence of
bits. The presented QKD protocols are provably secure against eavesdropping attack,
in that, as a matter of fundamental principle, the secret data can not be
compromised unknowingly to the legitimate users of the channel. Several 
quantum key distribution protocols have been proposed, all these protocols 
can be classed into two kinds. i) The point-to-point (two parties) quantum key distribution (pQKD). 
Three main protocols of these are the BB84 protocol [3], B92 protocol [4] and EPR 
protocol [5-7]. ii) The networking quantum key distribution (nQKD), e.g., the time-reserved EPR protocol [8]. The physical implementation may be refer to Townsend's works [13,14].

For the pQKD scheme, the first quantum key distribution scheme, i.e., the Bennett Brassard
(BB84) scheme, was presented a decade ago. It is implemented by the four states $\{|\uparrow\rangle, 
|\downarrow\rangle, |\nearrow\rangle, |\searrow\rangle \}$, where any of the two states $\{|\uparrow\rangle, |\downarrow\rangle\}$ and any of the two states $\{|\nearrow\rangle, |\searrow\rangle\}$ are non-commuted, 
$\{|\uparrow\rangle, |\downarrow\rangle\}$ may be any orthogonal states of two-dimensional 
Hilbert space. Its security is warranted by the uncertainty principle of quantum
mechanics. In 1992, Bennett devised another protocol, i.e., the B92 protocol, which is based 
on the transmission of nonorthogonal
quantum states. This protocol uses any two nonorthogonal states to implement the QKD. Its 
security relies on the no-cloning of unknown two-nonorthogonal states.
A further elegant scheme has been proposed by Ekert, which is implemented by the Einstein-Podolsky-Rosen (EPR) pairs [24]. It is called the EPR protocol
which relies on the violation of the Bell inequalities [25] to provide the secret security. 
Consider the two-particle correlation, Bennett {\sl et al} presented a modified version, in 
which the security is warranted by the quantum correlation.

Recently, Biham, Huttner, and Mor proposed a nQKD protocol, called time-reserved EPR protocol,
in which users prestore quantum states in a trusted center, where their quantum states are 
preserved using quantum memories. The main procedures are as follows: users store quantum 
states in quantum memories, kept in a transmission center. Upon request from two users, 
the center uses two-bit gates to project the product state of two noncorrelated particles 
(one from each user) onto a fully entangled state. As a result, the two users can share a 
secret bit, which is unknown even to the center. The time-reserved EPR protocol was proposed to
be used in a quantum cryptographic networking (QCN). The implementation of the time-reserved EPR scheme needs four 
particle for obtaining one qubit. One may ask that whether the nQKD scheme may be 
implemented by three or two particle or not? In this paper we propose several QKD protocols that use three or two particles to obtain one qubit. These protocols can also be used in networking QKD.

In this work we suggest two nQKD schemes. The suggested schemes need three 
parties: the trusted information center and two users, by conventional 
called Alice and Bob, here the center is trusted. One scheme is implemented by the 
Greenberger-Horne-Zeilinger (GHZ) [26] triplet state, we call this protocol as 
GHZ-nQKD protocol, in which the center's role is to measure his/her particle (from the GHZ triplet) by the random measurement like that in BB84 protocol, 
and tell Alice and Bob the measurement results.
Another is implemented by the Bell states, we call it as Bell-nQKD protocol. 
This scheme needs the center to measure the two-particle entanglement system by the Bell operators or the linear combination of Bell operators before 
Alice's and Bob's measurement and send the users his/her results.
By the center's assistance, the users can obtain the secret key,
the cheating center as well as the eavesdropper (Eve) can not eavesdrop the key. 

The paper is outlined as follows: First, in Sec. II, we review the  
two-particle maximally entangled states, the so-called Bell states and the 
three-particle maximally entangled states, the so-called GHZ triple state. In addition 
we investigate the correlation properties of the GHZ triplet and the Bell states. In Sec. III, we propose three protocols, which are implemented 
by the GHZ triplet. The efficiencies of these protocols are different, but they are more practical, especially the protocol 3. The securities of these protocols are analyzed.
In Sec. IV, we propose two protocols implemented by the Bell states. The securities 
of these protocols are investigated. In Sec. V, we discuss the applications of our protocols in network QKD. Conclusions are presented in Sec VI. 

\begin{flushleft}
{\bf II. The Bell states and the GHZ triplet states}
\end{flushleft}

First we review the two and three particles entanglement states.
In general, N-particle entanglement states may be written as [27]

\begin{equation}
|\psi\rangle=\prod^N_{i=1}|u_i\rangle\pm \prod^N_{i=1}|u^c_i\rangle,
\end{equation}
where $u_i$ stands for a binary variable $u_i\in \{|z+\rangle, |z-\rangle\}$ and $u^c_i=1-u_i$, 
$|z+\rangle$ and $|z-\rangle$ denote the spin eigenstates, or equivalently the horizontal and 
vertical polarization eigenstates, or equivalently any two-level system.
For $N=2$ they reduce to the Bell states and $N=3$ and $N=4$ they represent the GHZ 
states. For a general $N$ we shall calling them cat states. In this paper, we are 
interested in the case of $N=2$ and $N=3$, i.e., the Bell states and the GHZ triplet state.

\begin{flushleft}
{\bf 1. Bell states}
\end{flushleft}

Eq.(1) reduces to the Bell states when $N=2$

\begin{equation}
|\Psi^+\rangle_c=\frac{1}{\sqrt{2}}(|z+\rangle_a|z+\rangle_b+|z-\rangle_a|z-\rangle_b),
\end{equation}

\begin{equation}
|\Psi^-\rangle_c=\frac{1}{\sqrt{2}}(|z+\rangle_a|z+\rangle_b-|z-\rangle_a|z-\rangle_b),
\end{equation}

\begin{equation}
|\Phi^+\rangle_c=\frac{1}{\sqrt{2}}(|z+\rangle_a|z-\rangle_b+|z-\rangle_a|z+\rangle_b),
\end{equation}

\begin{equation}
|\Phi^-\rangle_c=\frac{1}{\sqrt{2}}(|z+\rangle_a|z-\rangle_b-|z-\rangle_a|z+\rangle_b),
\end{equation}
where the subscripts $c, a, b$ denote the states for the information center and the two communicators 
Alice and Bob. 
These Bell states can be generated from a type-II parametric down-conversion crystal [28]. Define the $x$ eigenstates

\begin{equation}
|x+\rangle=\frac{1}{\sqrt{2}}(|z+\rangle+|z-\rangle),
\end{equation}

\begin{equation}
|x-\rangle=\frac{1}{\sqrt{2}}(|z+\rangle-|z-\rangle),
\end{equation}
the four Bell states can be rewritten as

\begin{equation}
|\Psi^+\rangle_c=\frac{1}{\sqrt{2}}(|x+\rangle_a|x+\rangle_b+|x-\rangle_a|x-\rangle_b),
\end{equation}

\begin{equation}
|\Psi^-\rangle_c=\frac{1}{\sqrt{2}}(|x+\rangle_a|x-\rangle_b+|x-\rangle_a|x+\rangle_b),
\end{equation}

\begin{equation}
|\Phi^+\rangle_c=\frac{1}{\sqrt{2}}(|x+\rangle_a|x+\rangle_b-|x-\rangle_a|x-\rangle_b),
\end{equation}

\begin{equation}
|\Phi^-\rangle_c=\frac{1}{\sqrt{2}}(|x-\rangle_a|x+\rangle_b-|x+\rangle_a|x-\rangle_b),
\end{equation}

As should be noted, for example, the $|\Psi^+\rangle$ states give correlated results in 
both the $z$ and $x$ bases, but the $|\Psi^-\rangle$ state give correlated results in 
the $z$ basis, but anticorrelated results in the $x$ basis. Summarizing these correlated or 
anticorrelated results of the Bell states $\{\Psi^+, \Psi^-, \Phi^+, \Phi^-\}$ in the 
$z$ and $x$ bases, we get the following table:

\begin{center}
Table I. The correlation of Bell states $\{\Psi^+, \Psi^-, \Phi^+, \Phi^-\}$\\[0.1cm]

\begin{tabular}{c|llll}
\hline
\hline
Trent  &$|\Psi^+\rangle$ &$|\Psi^-\rangle$ &$|\Phi^+\rangle$ &$|\Phi^-\rangle$\\
\hline
Alice &$|x+\rangle$  &$|x+\rangle$  &$|x+\rangle$ &$|x-\rangle$ \\
Bob   &$|x+\rangle$  &$|x-\rangle$  &$|x+\rangle$ &$|x+\rangle$ \\
\hline
Alice &$|x-\rangle$  &$|x-\rangle$  &$|x-\rangle$ &$|x+\rangle$ \\
Bob   &$|x-\rangle$  &$|x+\rangle$  &$|x-\rangle$ &$|x-\rangle$ \\
\hline
Alice &$|z+\rangle$  &$|z+\rangle$  &$|z+\rangle$  &$|z+\rangle$ \\
Bob   &$|z+\rangle$  &$|z+\rangle$  &$|z-\rangle$  &$|z-\rangle$ \\
\hline
Alice &$|z-\rangle$  &$|z-\rangle$  &$|z-\rangle$  &$|z-\rangle$ \\
Bob   &$|z-\rangle$  &$|z-\rangle$  &$|z+\rangle$  &$|z+\rangle$ \\
\hline
\hline
\end{tabular}
\end{center}

From Table I it is clear that after the center has projected the two-particle 
entanglement system onto any of the four Bell states $\{\Psi^+, \Psi^-, \Phi^+, \Phi^-\}$, 
the state of any of two particles do not give determined results. For example, if the center's measurement basis is $|\Psi^+\rangle$, the state of any 
of two particles may be $|x+\rangle$,or $|x-\rangle$ with the probability $\frac{1}{2}$, or $|z+\rangle$ or $|z-\rangle$ with the probability $\frac{1}{2}$. 
Even if Alice has measured her particle and announced her measurement basis, anyone, including
Bob, can not knows Alice's results, because the probability of making error is 
$\frac{1}{2}$. However if the center's bases are public announced, Alice knows the 
Bob's qubits and vice versa. These properties may be used to distribute the quantum 
key between Alice and Bob by the assistance of the trusted center.

By the four Bell states (Eq.2-Eq.5) one may obtain other correlated or anticorrelated 
results. Define a line combination of Bell states as 

\begin{equation}
|\psi^+\rangle_c=\frac{1}{\sqrt{2}}(|\Psi^-\rangle_c+|\Phi^+\rangle_c),
\end{equation}
\begin{equation}
|\psi^-\rangle_c=\frac{1}{\sqrt{2}}(|\Psi^-\rangle_c-|\Phi^+\rangle_c).
\end{equation}
One may get
\begin{equation}
\begin{array}{rl}
|\psi^+\rangle_c=&\frac{1}{\sqrt{2}}(|x+\rangle_a|z+\rangle_b+|x-\rangle_a|z-\rangle_b)\\
               &\\
               &\frac{1}{\sqrt{2}}(|z+\rangle_a|x+\rangle_b+|z-\rangle_a|x-\rangle_b),
\end{array}
\end{equation}
\begin{equation}
\begin{array}{rl}
|\phi^-\rangle_c=&\frac{1}{\sqrt{2}}(|x+\rangle_a|z-\rangle_b-|x-\rangle_a|z+\rangle_b)\\
               &\\
               &\frac{1}{\sqrt{2}}(|z+\rangle_a|x-\rangle_b+|z-\rangle_a|x+\rangle_b),
\end{array}
\end{equation}
We note that the set of states $\{\Phi^+, \Psi^-, \phi^-, \psi^+\}$ have the 
following correlated or anticorrelated results 

\begin{center}
Table II. The correlation of states $\{\Phi^+, \Psi^-, \phi^-, \phi^+\}$\\[0.1cm]

\begin{tabular}{c|llll}
\hline
\hline
Trent  &$|\Phi^+\rangle$ &$|\Psi^-\rangle$ &$|\phi^-\rangle$ &$|\psi^+\rangle$\\
\hline
Alice &$|x+\rangle$  &$|x+\rangle$  &$|x+\rangle$ &$|x+\rangle$ \\
Bob   &$|x+\rangle$  &$|x-\rangle$  &$|z-\rangle$ &$|z+\rangle$ \\
\hline
Alice &$|x-\rangle$  &$|x-\rangle$  &$|x-\rangle$ &$|x-\rangle$ \\
Bob   &$|x-\rangle$  &$|x+\rangle$  &$|z+\rangle$ &$|z-\rangle$ \\
\hline
Alice &$|z+\rangle$  &$|z+\rangle$  &$|z+\rangle$  &$|z+\rangle$ \\
Bob   &$|z-\rangle$  &$|z+\rangle$  &$|x-\rangle$  &$|x+\rangle$ \\
\hline
Alice &$|z-\rangle$  &$|z-\rangle$  &$|z-\rangle$  &$|z-\rangle$ \\
Bob   &$|z+\rangle$  &$|z-\rangle$  &$|x+\rangle$  &$|x-\rangle$ \\
\hline
\hline
\end{tabular}
\end{center}

This table shows the states $\{\Phi^+, \Psi^-, \phi^-, \psi^+\}$ also have the correlation properties in $x$ and $z$ direction. If the center projects the two-particle entanglement system onto any of the four bases $\{\Phi^+, \Psi^-, \phi^-, \psi^+\}$, and sends respectively Alice and Bob one of two-particle entanglement, Alice's and Bob's particles have yet not a determined results before their measurement. 
For example, if the center measure the two-particle system using the
base $\phi^-$, Alice's measurement may be $|x+\rangle$ or $|x-\rangle$ if she measures her
particle use $x$ basis. Before Alice reveals her measurement bases, anyone can not know Alice's results, even if Bob. However, if Alice and Bob know the state measured by the center and their
measurement directions are determined, they can judge the qubits each other.

\begin{flushleft}
{\bf 2. GHZ triplet states}
\end{flushleft}

Eq.(1) reduces to eight GHZ triplet states for $N=3$. In this paper we use the following state

\begin{equation}
|\psi\rangle=\frac{1}{\sqrt{2}}(|z+z+z+\rangle+|z-z-z-\rangle).
\end{equation}

Suppose the center, Alice and Bob share one particle each from a three-particle entangled 
GHZ state, then the GHZ state may be represented by

\begin{equation}
|\psi\rangle_=\frac{1}{\sqrt{2}}(|z+\rangle_c|z+\rangle_a|z+\rangle_b+|z-\rangle_c
|z-\rangle_a|z-\rangle_b),
\end{equation}
where the first particle is that of the center, the second that of Alice, and the third that of Bob. Define the $y$ eigenstates

\begin{equation}
|y+\rangle=\frac{1}{\sqrt{2}}(|z+\rangle+i|z-\rangle),
\end{equation}

\begin{equation}
|y+\rangle=\frac{1}{\sqrt{2}}(|z+\rangle-i|z-\rangle),
\end{equation}
and using the $x$ eigenstates defined in Eq.(6,7), the GHZ triplet state can be 
rewritten as 

\begin{equation}
\begin{array}{rl}
|\psi\rangle=&\frac{1}{2}[(|x+\rangle|x+\rangle+|x-\rangle|x-\rangle)|x+\rangle\\
             & \\ 
             &+(|x+\rangle|x-\rangle+|x-\rangle|x+\rangle)|x-\rangle],
\end{array}
\end{equation}
or 
\begin{equation}
\begin{array}{rl}
|\psi\rangle=&\frac{1}{2}[(|y+\rangle|y-\rangle+|y-\rangle|y+\rangle)|x+\rangle\\
             & \\
             &+(|x+\rangle|x-\rangle+|x-\rangle|x+\rangle)|x-\rangle],
\end{array}
\end{equation}
or 
\begin{equation}
\begin{array}{rl}
|\psi\rangle=&\frac{1}{2}[(|y+\rangle|x-\rangle+|y-\rangle|x-\rangle)|y+\rangle\\
             & \\
             &+(|y+\rangle|x+\rangle+|y-\rangle|x-\rangle)|y-\rangle],
\end{array}
\end{equation}
or 
\begin{equation}
\begin{array}{rl}
|\psi\rangle=&\frac{1}{2}[(|x+\rangle|y-\rangle+|x-\rangle|y+\rangle)|y+\rangle\\
             & \\
             &+(|x+\rangle|y+\rangle+|x-\rangle|y-\rangle)|y-\rangle].
\end{array}
\end{equation}
The above decomposition demonstrates the correlation among three particles. For example, in 
Eq.(20) if one particle is in the state $|x+\rangle$ and the second particle is in the state
$|x+\rangle$, the third particle must be in the state $|x+\rangle$ because of the correlation 
of the GHZ triplet state. By Eqs.(20-23), one may construct a lock-up table to summarize these properties of GHZ states. 

\begin{center}
Table III. The correlation results of the GHZ triplet states\\[0.1cm] 

\begin{tabular}{c|llll}
\hline
\hline
Trent  &$|x+\rangle$ &$|x-\rangle$ &$|y+\rangle$ &$|y-\rangle$\\
\hline
Alice &$|x+\rangle$  &$|x+\rangle$  &$|x+\rangle$ &$|x+\rangle$ \\
Bob   &$|x+\rangle$  &$|x-\rangle$  &$|y-\rangle$ &$|y+\rangle$ \\
\hline
Alice &$|x-\rangle$  &$|x-\rangle$  &$|x-\rangle$ &$|x-\rangle$ \\
Bob   &$|x-\rangle$  &$|x+\rangle$  &$|y+\rangle$ &$|y-\rangle$ \\
\hline
Alice &$|y+\rangle$  &$|y+\rangle$  &$|y+\rangle$  &$|y+\rangle$ \\
Bob   &$|y-\rangle$  &$|y+\rangle$  &$|x-\rangle$  &$|x-\rangle$ \\
\hline
Alice &$|y-\rangle$  &$|y-\rangle$  &$|y-\rangle$  &$|y-\rangle$ \\
Bob   &$|y+\rangle$  &$|y-\rangle$  &$|x+\rangle$  &$|x-\rangle$ \\
\hline
\hline
\end{tabular}
\end{center}

The table III shows several properties of the GHZ triplet state: 
i) anyone of the three parties, i.e., the center, 
Alice or Bob, can determine whether the other two participators' results are the 
same or opposite and also that he (she) will gain no knowledge of what their 
results actually are, if he (she) knows what measurements have been made by 
the other two participators (that is $x$ or $y$). 
ii) From table III it is clear that allows two parties jointly, but only jointly, 
to determine which was the measurement outcome of the third party. So if 
the measurement directions of 
the three participators are public, the combined results of any two participators 
can determine what the result of the third party's measurement was.

\begin{flushleft}
{\bf III. GHZ-nQKD protocols}
\end{flushleft}

GHZ states has already found a number of uses. They form the basis of a very 
stringent test of local realistic theories. It was also proposed that they can 
be used for cryptographic conferencing or for multiparticle generations of 
superdense coding [27]. In addition, related states can be used to reduce 
communication complexity. Recently, it was proposed that they can be used 
for quantum secret sharing and quantum information split [19]. In this paper, we 
use the GHZ state to distribute quantum key between Alice and Bob by the 
center's assistance. In the following we present our quantum key distribution 
protocols.

\begin{flushleft}
{\bf A. The protocols}
\end{flushleft}

As discussed in Sec. II, the GHZ state has correlation properties that if only one communicator's measurement results is announced, the states of other two particles are still not determined, but two communicators' results can determine the third result. These properties may be used in the QKD relying on a third party.
Let us now show how to implement our quantum key distribution scheme by the GHZ state.  There
are several way to distribute the communicators the key by the center's assistance.

{\it {\bf Protocol 1}
\begin{enumerate}
\item The center measures his GHZ particle in the $x$ direction and obtains the result 
$|x+\rangle$ or $|x-\rangle$

\item The center tell Alice and Bob his measurement results.

\item Alice and Bob make respectively the random measurement on their GHZ particles, either in the $x$ or $y$ direction.

\item Check the eavesdropping by using the correlation properties of the GHZ states.

\item Alice and Bob compare their bases. If their measurement bases are same, Alice and Bob keep their results, otherwise they discard their results.

\item Alice and Bob obtain the final key by using the data sifting, the error 
correction and the privacy amplification technologies.

\end{enumerate}
}

In this protocol, we let the center firstly measures his particle from the GHZ triplet, 
and only measure it in the $x$ direction. The center's results will be $|x+\rangle$ or 
$|x-\rangle$. After the center has finished the measurement, Alice and Bob measure their 
particles. This protocol only uses the correlation results in the first and second columns 
of table III. Of course, the center may randomly measure his GHZ particles, either in the 
$x$ or $y$ direction, but the efficiency is low by this way, because these results 
measured along $y$ direction have no use in this protocol and a half particles will 
be discarded, the efficiency is only 12.5\%. The center's measurement collapses the
GHZ triplet state to be a two-particle system. The state of the two-particle entanglement is not 
determined, because they may be any of the states
$$|\Psi^1\rangle_{ab}=\frac{1}{\sqrt{2}}(|x+\rangle|x+\rangle+|x-\rangle|x-\rangle),$$
$$|\Psi^2\rangle_{ab}=\frac{1}{\sqrt{2}}(|x+\rangle|x-\rangle+|x-\rangle|x+\rangle),$$
$$|\Psi^3\rangle_{ab}=\frac{1}{\sqrt{2}}(|y+\rangle|y-\rangle+|y-\rangle|y+\rangle),$$
$$|\Psi^4\rangle_{ab}=\frac{1}{\sqrt{2}}(|y+\rangle|y+\rangle+|y-\rangle|y-\rangle).$$

In step 4, we use the correlation properties of the GHZ states to check the 
eavesdropping. Having measured their particles, Bob randomly chooses a subset of 
qubits from his qubits and sends this subset to Alice. Alice compares the 
corresponding results from the center, Bob and Alice. If these results are correlation 
results, which are satisfy Eqs.(20-23) or the correlation of three states is in the 
table III, the results are perfect, otherwise it means eavesdropping or disturbed
by noise.

In step 5, Alice and Bob compare their bases. Because Alice and Bob randomly measure their
particles either in the $x$ or $y$ direction, some of their bases are different and some are 
same. If their bases are different, Alice's and Bob's results are no correlation, thus Alice 
can not know Bob's qubits and vice versa, in this case they need to discard these results. However if their bases are same, their results are correlated, Alice and Bob keep these results. So this protocol discards the results that corresponds the different bases.

The raw quantum key distribution is useless in practice
because limited eavesdropping may be undetectable, yet it may leak some information, 
and errors are to be expected even in the absence of eavesdropping. For these reasons, 
our scheme needs to supplement some classical tools such as the privacy amplification, the error correction and the data sifting, so we use these technologies in our protocol. 
The implementation of these supplemented classic tools are the same as in the 
previous documents [9].

In quantum key distribution some qubits (henceforth $l$) will be wasted because of the 
loss and the inexactitude of equipment, so in order to be left with a key 
of $L$ qubits the center {\footnote {assume the particles are prepared by the center in this paper}} should prepare $L'>2(L+l)$. In this case the efficiency is
\begin{equation}
\eta_1 =\frac{L}{2(L+l)}<50\%.
\end{equation}
This efficiency is larger than that of the time-reserved EPR protocol, which is 
\begin{equation}
\eta' =\frac{L}{8(L+l)}<12.5\%.
\end{equation}

{\it {\bf Protocol 2}
\begin{enumerate}
\item The center measures his GHZ particle either in the $x$ or $y$ direction and obtains any of the four states $\{|x+\rangle, |x-\rangle, |y+\rangle, |y-\rangle\}$

\item The center tells Alice and Bob his measurement results.

\item Alice and Bob make respectively a random measurement on their GHZ particles, either in the $x$ or $y$ direction.

\item Checking the eavesdropping like protocol 1.

\item Alice and Bob compare their bases. If the center's result is $|x+\rangle$ or 
$|x-\rangle$ and their measurement bases are same, or if the center's result is $|y+\rangle$ 
or $|y-\rangle$ and their measurement bases are different, Alice and Bob keep their results,
otherwise they discard the results.

\item Alice makes her results to be consistent with Bob's results according to table III.

\item Alice and Bob gain the final key by using the data sifting, the error 
correction and privacy amplification technologies.

\end{enumerate}
}

The protocol 2 lets the center measure his/her particle either in the $x$ or $y$ direction. 
It is stresses that here the center's all measurement results are useful. When the center's result 
is the state $|x+\rangle$ or $|x-\rangle$, Alice and Bob need to keep the results which have 
the same bases, but if the center's result is $|y+\rangle$ or $|y-\rangle$, the communicators 
discard the results which have the same bases. The reason is that the results must be 
correlated or anticorrelated. This step is finshed in the step 5.

It needs to stress that Alice's results must be consistent with Bob's results for getting the raw quantum key, so we have the step 6. By the properties of the GHZ triplet state, Alice (Bob) can judge 
Bob's (Alice's) results by combining her (his) and the center's results. 
But the table III can not give completely a same results although Alice and Bob can 
know the qubits each other. For example, when the center's and Alice's results are 
respectively $|y+\rangle$ and $|x+\rangle$, Bob's result should be $|y-\rangle$, obviously,
Alice's and Bob's results are different.
For obtaining a same key, Alice's (Bob's) results need to be consistent with Bob's 
(Alice's) results. The method is that Alice (Bob) transfers her (his) qubits to binary 
bits according to Bob's (Alice's) results.

The efficiency of this protocol is the same as the protocol 1. After the center announced his
results, Alice and Bob have a possibility of $1/2$ to obtain the correct results by the 
random measurement. Consider the wasted qubits ($l$), in order to be left with a key 
of $L$ qubits the center should send $L'>2(L+l)$, the efficiency 
\begin{equation}
\eta_2 =\frac{L}{2(L+l)}<50\%.
\end{equation}

{\it {\bf Protocol 3}
\begin{enumerate}
\item Alice and Bob make respectively a random measurement on their GHZ particles, 
either in the $x$ or $y$ direction. 

\item Alice and Bob send their measurement bases ($x$ or $y$) to the center, but not 
the qubit values.

\item The center randomly measures his particle according to Alice's and Bob's 
measurement bases. 
If both Alice and Bob measure their particle using the same measurement basis, 
e.g. $x$ or $y$ direction, the center measures his particle using the $x$ 
measurement basis, otherwise, the center measures his particle using the $y$ 
measurement basis. 

\item The center announces his measurement results, which is any of the four states 
$\{|x+\rangle, |x-\rangle, |y+\rangle, |y-\rangle\}$. 

\item Check eavesdropping like the protocol 1.

\item Alice and Bob judge the quantum state each other according to table III. 
While the center announces his results, both Alice and Bob know the center's 
results. Then the Alice's (Bob's) 
and the center's results can jointly determine what is the Bob's (Alice') measurement outcome. 

\item Alice and Bob obtain a sharing key by using the data sifting, the error 
correction and privacy amplification technologies.

\end{enumerate}
}

The important point is that Alice and Bob randomly measure their GHZ particles before 
the center's measurement in this protocol, it is different from the protocols 1 and
2, in which the center's measurement is completed before Alice's and Bob's 
measurement. This change improves the efficiency of this protocol, because the center 
may measure his particle according to Alice's and Bob's measurement bases. Although there are the situations that Alice's and Bob's results are different, all their measurement states are useful. The correlation of the 
GHZ triplet state lets Alice know Bob's qubits and vice versa, so Alice and Bob can know the qubits each other in the step 6. Consider 
the wasted qubits ($l$) in the measurement and the loss in the quantum channel, in 
order to be left with a key 
of $L$ qubits the center should send $L'>(L+l)$, the efficiency is
\begin{equation}
\eta_3 =\frac{L}{(L+l)}<100\%.
\end{equation}

\begin{flushleft}
{\bf B. security analysis}
\end{flushleft}

These presented schemes are secure against eavesdropping. Their securities are 
warranted by the correlation of the GHZ triplet. To see these in a sufficient way, we will consider several possible eavesdropping in the following. 

\begin{flushleft}
{\bf 1. The cheating center's attacks}
\end{flushleft}

The cheating center is impossible to know the quantum key. From the table III it is
clear that if the center knows what measurements bases Alice and Bob made (that is, $x$ or $y$), he
can determine whether their results are the same or opposite and also that 
the center will gain no knowledge of what Alice's and Bob's actually are, because the cheating
center will has the probability of 1/2 of making a mistake. If the center makes measurement on the three particles of the GHZ triplet, then send these particles to Alice and Bob, the center's measurement will
introduce errors in Alice's and Bob's results, thus Alice and Bob can check it like the
four-state BB84 protocol. Of course, a cheating center may use the men-in-middle attack [29] to 
obtain the key $K_{ca}$ and $K_{cb}$, where $K_{ca}$ represents the key between Alice and
the cheating center, and  $K_{cb}$ represents the key between Bob and the cheating center. 
For preventing this attacks, Alice and Bob may verify their identity using the identity verification technology [30]. This method needs a sharing key between Alice and Bob, however, Alice and 
Bob, in general, have not sharing key, so this case needs the center to be trustworthy
like the key distribution system used in the 
classic cryptography [29]. So in our scheme we assume the center is trusted, e.g., the key
distribution center (KDC) which is often used in the classic cryptography, but here the center
process the qubits not the binary bits.

\begin{flushleft}
{\bf 2. Intercept/resend attacks}
\end{flushleft}

Let us now consider the intercept/resend attack defined in [9]. Suppose that the 
eavesdropper, by convention denoted by Eve, has managed to get a hold of Alice and Bob's key, 
she then intercepts a communicator's (e.g. Alice) particle from the center and send another particle to Alice. In this
case, three particles of the center, Bob and the Eve construct a GHZ triplet. However, because the Alice, Bob and the center's particles are
not the GHZ triplet, there are no correlated or anticorrelated result, Eve's interception will 
introduce error and can be detected by Alice and Bob when they check the eavesdropping. 

\begin{flushleft}
{\bf 3. The entanglement attacks}
\end{flushleft}

The entanglement attacks is no use in our protocol. To show that, Let us assume that the eavesdropper has been able to entangle an ancilla in state $|A\rangle$ with the GHZ
triplet state that Alice and Bob are using. The state describing the state of the GHZ 
triplet and the ancilla is 
\begin{equation}
|\Psi\rangle=\frac{1}{\sqrt{2}}(|z+z+z+\rangle+|z-z-z-\rangle)\otimes|A\rangle.
\end{equation}
By using the $x$ and $y$ eigenstates and Eq.(20), The eavesdropper get

\begin{equation}
\begin{array}{rl}
U|\Psi\rangle=&\frac{1}{2}(|x+x+\rangle\otimes |x+\rangle\otimes |A_1\rangle
+|x-x-\rangle\otimes |x+\rangle\otimes |A_2\rangle\\
&\\
&+|x+x-\rangle\otimes |x-\rangle\otimes |A_3\rangle
+|x-x+\rangle\otimes |x-\rangle\otimes |A_4\rangle
).
\end{array}
\end{equation}
where $U$ denote the unitary transformation. By projecting the above state onto 
$|\phi\rangle=\alpha_1|x+x+\rangle+\alpha_2|x-x-\rangle+\alpha_3|x+x-\rangle
+\alpha_4|x-x+\rangle$, the eavesdropper creates the states
\begin{equation}
|\Psi\rangle_E=\frac{1}{2}
(|x+\rangle\otimes (\alpha^*_1|A_1\rangle+\alpha^*_2|A_2\rangle)
+|x-\rangle\otimes (\alpha^*_3|A_3\rangle+\alpha^*_4|A_4\rangle)
).
\end{equation}
If Eve can gain Alice's (Bob's) qubits, she can obtain the key. However,
Eq.(30) shows that Eve can not obtain the results. By the similar method, Eve can 
not obtain Alice's (Bob's) qubits when the GHZ triplet states  satisfy Eqs.(21-23).

\begin{flushleft}
{\bf IV. Bell-nQKD protocols}
\end{flushleft}

The above schemes are efficient, however they need three particles. Can we implement the network QKD scheme only by using
two particles? In this section we investigate the two-particle schemes. In the following
we show that one can also use the Bell states to implement the above quantum key distribution procedure. 

\begin{flushleft}
{\bf A. protocol}
\end{flushleft}

In Sec. II, we see that the two particles of the Bell states or the linear combination of 
Bell states have correlation properties, they are demonstrated in table I and II.
These properties may be used in the QKD relied on a third party. 
Let us now show how to implement the quantum key distribution by Bell states.  

{\it {\bf Protocol 4}
\begin{enumerate}
\item The center prepares a set of two-particle entanglement pairs and projects each pair onto any of the four Bell bases.

\item The center sends respectively Alice and Bob one of the two-particles entanglement 
and his measurement results.

\item Alice and Bob make respectively the random measurement on their particle, either 
in the $x$ or $z$ direction.

\item Alice and Bob check the eavesdropping by using the correlation of Bell states.

\item If their measurement bases are same, Alice and Bob keep their results, otherwise they discard the results.

\item Alice and Bob obtain a sharing key by using the data sifting, the error 
correction and the privacy amplification technologies.

\end{enumerate}
}

This scheme is similar to the time-reserved EPR protocol, but there are several important 
dissimilarities. i) The time-reserved EPR protocol uses four particle and two particles 
were prestored in a transmission center, where their quantum states are preserved using 
quantum memories. Our scheme uses two particles and need not the quantum 
memories. ii) The efficiency of the time-reserved EPR protocol is 
$\eta'<12.5\%$, but the efficiency of our protocol is 
\begin{equation}
\eta_4=\frac{L}{2(L+l)}<50\%. 
\end{equation}
iii) The center only uses results of the singlet states and its 
correlation properties in the time-reserved EPR protocol, but the center uses all 
quantum states in our scheme.

We can also use the table II to design a nQKD protocol. The protocol goes as
follows

{\it {\bf Protocol 5}
\begin{enumerate}
\item The center prepares a set of two-particle entanglement pairs and projects each pair onto any of the four bases $\{\Phi^+, \Psi^-, \phi^-, \psi^+,\}$.

\item The center sends respectively Alice and Bob one of the two-particles entanglement 
and his measurement results.

\item Alice and Bob make respectively the random measurement on their particle, either 
in the $x$ or $z$ direction.

\item Check the eavesdropping by using the correlation demonstrated in Table II.

\item Alice and Bob compare their bases. If the center's result is one of the Bell states $\{|\Phi^+\rangle, |\Psi^-\rangle\}$ and their measurement bases are same, or if the 
center's result is one of the states $\{|\psi^+\rangle, |\phi^-\rangle\}$ and 
their measurement bases are different, Alice and Bob keep their results,
otherwise they discard the results.

\item Alice makes her results be consistent with Bob's results.

\item Alice and Bob obtain a sharing key by using the data sifting, the error 
correction and privacy amplification technologies.

\end{enumerate}
}

This protocol is similar to the protocol 2, but there are two dissimilarities: 1) The implementation of protocol 2 uses 
the GHZ triplet state, which needs three particles to obtain one qubit. The center's measurement result is one of the state $\{|x+\rangle,
|x-\rangle, |y+\rangle, |y-\rangle\}$. But the  protocol 5 uses the Bell states which 
only use two particles, the center's results is one of the states $\{|\Phi^+\rangle,
|\Psi^-\rangle, |\psi^+\rangle, |\phi^-\rangle\}$. 2) The methods for checking eavesdropping
are different. Protocol 2 uses the correlation of the GHZ states, and here protocol 5 uses the correlation demonstrated in the table II.

According to the protocol 5, we see the efficiency is same as protocol 2:
\begin{equation}
\eta_5=\frac{L}{2(L+l)}<50\%. 
\end{equation}

\begin{flushleft}
{\bf B. security analysis}
\end{flushleft}

In term of eavesdropping possibilities, protocols 4 and 5 have same security with the 
EPR protocol. After the center has measured the two-particle entanglement systems by 
using any Bell operators or the linear combination Bell operators, Alice and Bob's 
particles are two-particle entanglement pairs, which is one of the four states 
$\{|\Psi^+\rangle, |\Psi^-\rangle, |\Phi^+\rangle, |\Phi^-\rangle,\}$ or 
$\{|\Phi^+\rangle, |\Psi^-\rangle, |\phi^-\rangle, |\psi^+\rangle,\}$. It has the same
correlation as the EPR pair, this is therefore equivalent to the EPR scheme. 
So the cheating center as well as the eavesdropper can not eavesdrop the key from
the protocols 4 and 5 by the currently eavesdropping technologies, e.g., the 
intercept/resend attacks, the entanglement attacks etc..

\begin{flushleft}
{\bf V. Practical applications}
\end{flushleft}

Like the time-reserved EPR protocol, ours scheme can also be used in the quantum 
cryptographic network (QCN), not only in the quantum cryptographic network with a centers 
and multiusers but also in the QCN of worldwide network of many center. For the QCN of worldwide network of many centers, we can also implement the network 
QKD between two users by the teleporation scheme.
The ways are similar to the time-reserved EPR protocol. As an example, we 
consider the QCN with a center and multiusers. Assume that there are $N$ users denoted by $u_i, i=1,2,\cdots, N$ and a trusted information center, all users have registered their
$ID$ in this system at the initial phase. When any user $u_i$ wants to connect the other 
user $u_j$ who is in the same QCN, the user $u_i$ first tells the center this message. The center 
verifies the user's identity by the quantum authentication technology [31] or the classic authentication scheme [32]. If the identity is correct, 
the center distributes GHZ particles or the two particles of Bell states to users $u_i$ and 
$u_j$, then they use the proposed protocol to distribute the key between two users. Here the users $u_i$ and $u_j$ are arbitrarily chosen. 

\begin{flushleft}
{\bf VI. conclusion}
\end{flushleft}

We have show that the quantum correlation between two particles and among three 
particles can be used for the quantum key distribution relying on a trusted 
information center. Two schemes are proposed. One scheme is implemented by using the 
GHZ triplet states, in which three protocols are proposed, these protocols use three 
particles to obtain one qubit, and have optimal efficiency, especially the protocol 
3. The other is implemented by the Bell state, in which two protocols are proposed, 
these protocols use one Bell particle pairs to obtain one qubit.

In the process of quantum key distribution, the center play
an important role, however, the center can not gain any information by any method except for the classic attack, i.e., the men-in-middle attack. For preventing the men-in-middle attack, the presented schemes need the trusted information center, or the users can verify the communicators' identity (in fact all
previous QKD protocols, e.g., BB84, B92 and EPR protocol need this requisition in practical application). 

\begin{flushleft}
{\bf Acknowledgments}
\end{flushleft}

This research was supported by the Natural Science Foundation (NSF) of China under 
the grants of No. 69803008

\begin{flushleft}
References
\end{flushleft}

\begin{enumerate}
\item S. Wiesner, Sigact News, {\bf 15}, 78 (1983); original manuscript written circa 1970. 

\item C. H. Bennett, G. Brassard, S. Breidbart, and S. Wiesner, Advances in Cryptology: 
Proceedings of Crypto 82, August 1982, Plenum Press, New York, p. 267. 

\item C. H. Bennett, and G.Brassard,  Advances in 
Cryptology: Proceedings of Crypto 84, August 1984, Springer - Verlag, p. 475. 

\item C. H.Bennett,  Phys. Rev. Lett., {\bf 68}, 3121, (1992). 

\item A. K.Ekert, Phys. Rev. Lett., {\bf 67}, 661, (1991).

\item A. K. Ekert, J. G. Rarity, P. R. Tapster, and G. M. Palma, 
Phys. Rev. Lett. {\bf 69}, 1293 (1992). 

\item C. H. Bennett, G. Brassard, and N. D. Mermin, Phys. Rev. Lett. {\bf 68}, 557 (1992).

\item E. Biham, B. Huttner, and T. Mor, Phys. Rev. A, {\bf 54}, 2651, (1996).

\item C. H. Bennett, F. Bessette, G. Brassard, L. Salvail and 
J. Smolin, J. Cryptology {\bf 5}, 3 (1992). 

\item A. Muller, J. Breguet, and N. Gisin, Europhysics Lett., vol. 23, no. 6, 
383 (1993). J. Breguet, A. Muller, and N. Gisin, J. Modern Optics, {\bf 41}, 2405, (1994).
\item C. Marand and P. D. Townsend, Optics Letters, {\bf 20}, 1695 (1995).

\item W. T. Buttler, R. J. Hughes, P. G. Kwiat, et. al., Phys. Rev. A {\bf 57}, 2379 (1998).

\item P. D. Townsend, J. G. Rarity, and P. R. Tapster, Electronics Letters, {\bf 29},  
634, (1993).
\item P. D. Townsend, J. G. Rarity, and P. R. Tapster, Electronics Letters, 
{\bf 29}, 1291, (1993).. 

\item C. A. Fuchs, N. Gisin, R. B. Griffiths, C. S. Niu, and A. Peres, 
Phys. Rev. A, {\bf 56}, 1163, (1997).

\item B. A. Slutsky, R. Rao, P. C. Sun, and Y. Fainman, 
Phys. Rev. A, {\bf 57}, 2383, (1998).

\item Brandt, Howard E., John M. Meyers, And Samuel J.Lomonaco,Jr., Phys. Rev. A, {\bf 56}, 4456, (1997).

\item C. Niu, and R. Griffiths, Phys. Rev. A, {\bf 58}, 4377, (1998).

\item M. Hillery, V. Buzek, and A. Berthiaume, Phys. Rev. A, {\bf 59}, 1829 (1999).
\item A. Karlsson, M. Koashi, and N. Imoto, Phys. Rev. A, {\bf 59}, 162 (1999). 
\item D. Mayers, Phys. Rev. Lett., {\bf 78}, 3414 (1997).
\item H. K. Lo and H. F. Chau, Phys. Rev. Lett., {\bf 78}, 3410 (1997).
\item L. Goldenberg, L. Vaidman, and S. Wiesner, Phys. Rev. Lett., {\bf 82}, 3356 (1999).
\item J. S. Bell, Physics (Long Island City, N.Y.), {\bf 1}, 195(1965).
\item A. Einstein, B. Podolsky, and N. Rosen, Phys. Rev. {\bf 47}, 777 (1935).

\item D. Greenberger, M. A. Horne, and A. Zeilinger, in Bell's Theorem, Quantum theory, and Conceptions of Universe, edited by M.Kaftos (Kluwer Academic, Dordrecht, 1989); 
D. Greenberger, M. A. Horne, A. Shimony, and A. Zeilinger, Am. J. Phys. 
{\bf 58}, 1131 (1990).

\item S. Bose, V. Vedral, and P. L. Knight, Phys. Rev. A, {\bf 57}, 822 (1998).

\item P. G. Kwiat, K. Mattle, H. Weinfurther, and A. Zeilinger, Phys. Rev. Lett., {\bf 75}, 
4337 (1995).

\item B. Schneier, Applied cryptography---Protocols, Algorithms, and Source Code in C, John Wiley \& Son, Inc., 1996.

\item G. Zeng and W. Zhang, Phys. Rev. A, {\bf 61} no 1, 2000 (In press).
\item G. Zeng, Phys. Lett. A, (Submitted).
\item M. N. Wegman, and J. L. Carter, J. Computer and System Sci., vol. 22, 265 (1981).
\end{enumerate}

\end{document}